\documentclass[11pt]{article}
\usepackage{amsmath,amssymb,graphicx}
\def\lp{\left(}
\def\rp{\right)}
\def\lb{\left[}
\def\rb{\right]}
\def\be{\begin{equation}}
\def\ee{\end{equation}}
\title{More about thin-shell wormholes associated to cosmic strings}
\author{Mart\'{\i}n G. Richarte\thanks{E-mail: martin@df.uba.ar}\ \ and Claudio Simeone\thanks{E-mail: csimeone@df.uba.ar. }\\
{\small  Departamento de F\'{\i}sica, Facultad de Ciencias Exactas y 
Naturales, UBA}
   \\ 
{\small Ciudad Universitaria, Pabell\' on I, 1428, 
Buenos Aires, Argentina}} 

\begin{document}
\maketitle

\begin{abstract}

\noindent Previous analysis about thin-shell wormholes associated to cosmic strings are extended.  More evidence is found supporting the conjecture that, under reasonable assumptions about the equations of state of matter on the shell, the  configurations are not stable under radial velocity perturbations.

\end{abstract}

\bigskip
\section{Introduction}
 
\noindent 

Following the leading work by Morris and Thorne \cite{mo}, traversable Lorentzian wormholes received considerable attention \cite{visser}. Wormholes would imply a nontrivial topology, connecting two regions of the universe by a traversable throat. If they actually exist, such configurations could include some features of unusual interest, as for example the possibility of time travel \cite{mo-fro}. However, the flare-out condition to be fulfilled at the wormhole throat requires --in the framework of General Relativity-- the existence of matter which violates the energy conditions (``exotic matter'') \cite{mo,visser,hovis}. Conditions reducing matter supporting wormhole geometries, consequently, have  deserved a detailed analysis. Besides (as most solutions of the equations of gravitation), traversable wormhole geometries result of physical interest mainly as long as  they are stable, at least under a simple kind of perturbation. Thus, beyond the mere characterization of static wormhole solutions, their stability under perturbations should always be explored.

Some years ago, the gravitational effects of topological deffects as domain walls and cosmic strings were the object of a thorough study,  because of their possible relevance in structure formation in the early universe, and also for their possible manifestation as gravitational lenses (see Ref. \cite{vilenkin}). On the other hand,  present theoretical developments suggest a scenario in which the fundamental building blocks of nature are extended objects. In particular, one-dimensional objects --strings-- constitute the candidates more often considered. Thus the interest in the gravitational effects of both cosmic and fundamental strings has been renewed in the last years (see, for example, Refs. \cite{strings}). Recent studies regarding cylindrical wormholes include the works by Cl\'ement \cite{cle1,cle2}, Aros and Zamorano \cite{arza}, Kuhfittig \cite{ku}, Bronnikov and Lemos \cite{brle} and by Bejarano, Eiroa and one of us \cite{ei1,ei2}. The  thin-shell wormhole configurations associated to local and global cosmic strings  analyzed in Refs. \cite{ei1,ei2} turned to be unstable under velocity perturbations preserving the cylindrical symmetry. Moreover, it was noted \cite{ei2} that this feature seemed to be independent of the particular geometry considered, as long as the symmetry and the form of the equations of state of the static configurations were preserved. Because for a far observer these wormholes would appear as cosmic strings, bounds on cosmic string abundance, which seem rather restrictive  (see for instance Ref. \cite{lar-his}), could imply a first constraint on the existence of such wormholes. Mechanical unstability would mean an additional restriction to the possibility of finding these objects in the present day universe.

In the present work we perform two extensions of these previous analysis regarding  cylindrical thin-shell wormholes associated to cosmic strings. We first construct thin-shell wormholes associated with two different global strings;
then, starting from static wormhole configurations, we consider small  velocity perturbations and determine the character of  the subsequent
dynamical evolution. Our results are consistent with the conjecture introduced in Ref. \cite{ei2}.

\section{Thin--shell wormholes}

The wormhole construction follows  the usual steps of the cut  and paste procedure \cite{musvis}. Starting from a manifold $M$
described by coordinates $X^\alpha=(t,r,\phi,z)$ we remove the region  defined by $r<a$ and take two copies $M^+$ and $M^-$ of
the resulting manifold. Then we join them at the surface $\Sigma$  defined by $r=a$, so that a new geodesically complete
manifold ${\cal M}=M^+\cup M^-$ is obtained. The surface $\Sigma$ is a  minimal area hypersurface satisfying the flare-out
condition: in both sides of the new manifold, surfaces of constant $r$  increase their areas as one moves away from $\Sigma$;
thus one says that ${\cal M}$ presents a throat at $r=a$. On the  surface $\Sigma$ we define coordinates $\xi=(\tau,\phi,z$), where
$\tau$ is the proper time. Then to allow for a dynamical analysis, we  let the radius depend on $\tau$, so that  the
surface $\Sigma$ is given by the function ${\cal H}$ which fulfills the  condition ${\cal H}(r,\tau)=r-a(\tau)=0$.
As a result of pasting the two copies of the original manifold, we  have a matter shell placed at  $r=a$. Its
dynamical evolution is determined by the Einstein equations projected  on $\Sigma$, that is, by the Lanczos equations \cite{lanczos}
\be
-[K_i^j]+[K]\delta_i^j=
8\pi S_i^j,
\label{e10}
\ee
where $K_i^j$ is the extrinsic curvature tensor defined by
\be
  K_{ij}^{\pm} = - n_{\gamma}^{\pm} \left. \left( \frac{\partial^2
  X^{\gamma}}{\partial \xi^i \partial \xi^j} + \Gamma_{\alpha \beta}^{\gamma}
  \frac{\partial X^{\alpha}}{\partial \xi^i} \frac{\partial
  X^{\beta}}{\partial \xi^j} \right) \right|_{\Sigma}, \label{e6}
\ee
with $n_{\gamma}^{\pm}$  the unit normals ($n^{\gamma} n_{\gamma} = 1$) to
$\Sigma$ in $\mathcal{M}$:
\be
  n_{\gamma}^{\pm} = \pm \left| g^{\alpha \beta} \frac{\partial
  \mathcal{H}}{\partial X^{\alpha}} \frac{\partial \mathcal{H}}{\partial
  X^{\beta}} \right|^{- 1 / 2} \frac{\partial \mathcal{H}}{\partial
  X^{\gamma}} . \label{e7}
\ee
The bracket  $[K_i^j]$ denotes the jump ${K_i^j}^+ - {K_i^j}^-$ across the surface $\Sigma$, 
$[K]=g^{ij}[K_{ij}]$ is the 
trace of $[K_{ij}]$ and
$S_i^j = {\rm diag} ( -\sigma, 
p_{\theta}, p_{z} )$ is the surface stress-energy tensor, 
with $\sigma$ the surface energy density and $p_\theta$, $p_z$  the 
surface pressures.
\section{Wormholes associated to cosmic strings}

\subsection{Nonsingular global string}

We first construct a thin-shell wormhole associated to a global cosmic  string whose metric, differing from the case treated in \cite{ei2}, presents  no curvature singularity; this is a consequence of relaxing the demand of boost invariance along the axis of symmetry \cite{senba}. The geometry has the form 
\be
ds^2= F(r)\lp -dt^2+dr^2\rp+H(r)dz^2+G(r)d\theta^2,\label{sen}
\ee
where
\begin{eqnarray}
F(r) & = & \lb P+\frac{2}{B^2}\ln \frac{r}{r_0}\rb^{1-\sqrt{2}},\\
H(r) & = & \lb P+\frac{2}{B^2}\ln \frac{r}{r_0}\rb^{\sqrt{2}},\\
G(r) & = & 4B^2\beta^2\lp\frac{r}{r_0}\rp^2\lb P+\frac{2}{B^2}\ln \frac{r}{r_0}\rb^{2-\sqrt{2}}.
\end{eqnarray}
The constant $r_0$ is the core radius, and $P$, $B$ and $\beta$ are integration constants fulfilling $B^2=4\pi v^2/\beta^2$, with $v$ defining the value of the field associated to the string \cite{senba}.  It is clear from the metric (\ref{sen}) that for $P>0$, which corresponds to the case of no outer singularity, the area per unir length is an increasing function of the radial coordinate for  $r>r_0$. Then the flare-out condition required for the wormhole construction is fulfilled. Also, note that the geodesics within a plane orthogonal to the string open up at the throat of the wormhole construction, because the metric component $g_{\theta\theta}(r)$ is itself an increasing function of $r$.

In terms of these functions the components of the extrinsic curvature read 
\be
{K_\tau^\tau}^{\pm} = \mp \frac{\sqrt{F(a)}}{2 \sqrt{1+F(a)\dot{a}^2}}\left( 2\ddot a+\frac{F'(a)}{F^2(a)}+2\dot a^2\frac{F'(a)}{F(a)}\right)
\label{e8a}
\ee
\be
{K_\theta^\theta}^{\pm} = \pm \frac{G'(a)\sqrt{1+F(a)
\dot{a}^2}}{2G(a)\sqrt{F(a)}}, 
\label{e8b}
\ee
and
\be
{K_z^z}^{\pm} = \pm \frac{H'(a)\sqrt{1+F(a)\dot{a}^2}}
{2H(a) \sqrt{F(a)}}, 
\label{e8c}
\ee
where the dot means $d/d\tau$ and the prime indicates a derivative with respect to $r$.
Replacing these expressions in the Lanczos equations we obtain the surface energy density $\sigma=-S_\tau^\tau$ and the pressures $p_\theta=S_\theta^\theta$ and $p_z=S_z^z$:  
\be
  \sigma = - \frac{\sqrt{1 + F ( a ) \dot{a}^2}}{8 \pi \sqrt{F ( a )}} \left( 
  \frac{G' ( a )}{G ( a )} + \frac{H' ( a )}{H ( a )} \right ),
\label{e11}
\ee
\be
p_{\theta} =  \frac{1}{8 \pi \sqrt{F(a)} \sqrt{1 + F(a)\dot{a}^2}} 
\left[2 F(a) \ddot{a} + F (a) \dot{a}^2 \left( \frac{H'(a)}{H(a)} + \frac{2 F'(a)}
{F(a)} \right )  + \frac{H'(a)}{H(a)} + \frac{F'(a)}{F(a)} \right] ,
\label{e12}
\ee
\be
p_{z} =  \frac{1}{8 \pi \sqrt{F ( a )} \sqrt{1 + F(a) \dot{a}^2}}
\left[ 2 F(a) \ddot{a} + F(a) \dot{a}^2 \left( \frac{G'(a)}{G(a)} + \frac{2 F'(a)}
{F(a)} \right) + \frac{G'(a)}{G(a)} + \frac{F'(a)}{F(a)} \right] .
\label{e13}
\ee
From these equations we find that for the static situation $\dot a=\ddot a=0$ the pressures and the surface energy density satisfy the equations of state
\be
p_\theta=-\sigma\frac{G(a)\lb F(a)H'(a)+F'(a)H(a)\rb}{F(a)\lb G(a)H'(a)+G'(a)H(a)\rb},
\ee
\be
p_z=-\sigma\frac{H(a)\lb F(a)G'(a)+F'(a)G(a)\rb}{F(a)\lb G(a)H'(a)+G'(a)H(a)\rb}.
\ee 
If we are interested in small velocity perturbations,  it is licit to assume that the evolution of the matter on the shell can be described as a succesion of static states. Thus we shall accept, as done before  \cite{ei1,ei2}, that the form of the equations of state corresponding to  the static case is kept valid in the dynamical evolution \footnote{Besides, the perturbative treatement avoids possible subtleties regarding the validity of the static geometry for $r>a$ in the presence of a cylindrical moving shell.}. With this assumption, the equations above lead to the equation of motion
\be
2F(a)\ddot a+F'(a)\dot a^2=0.
\ee
By writing $\ddot a=\dot a d\dot a/da$ we can recast this equation (for $\dot a\neq 0$) as
\be
2F(a)\frac{d\dot a}{da}+F'(a)\dot a=0.
\ee
This equation is solved by 
\be
\dot a(\tau)=\dot a_0\sqrt{\frac{F(a_0)}{F(a)}},
\ee 
where $a_0$ and $\dot a_0$ are, respectively, the initial wormhole radius and its inicial velocity. This shows that the sign of the velocity is determined by its initial sign, that is, after a small velocity perturbation the throat undergoes a monotonous evolution; no oscillatory behaviour exists, at least under the approximations assumed. Note that, because $F$ is a decreasing function of $a$, then the absolute value of the throat velocity decreases when $\dot a_0$ is negative, while the velocity grows when $\dot a_0$ points outwards (so, after a finite time the assumption of a low velocity is no more valid).

\subsection{Global string with curvature singularity}

After studying the nonsingular metric case, we now address a  generalization of the analysis carried out in Ref. \cite{ei2}.
 The starting point is the metric \cite{coka}
\be
ds^2=-U(r)dt^2+W(r)dz^2+V(r)\lp dr^2+r^2d\theta^2\rp\label{ckw}
\ee
with the functions $U$, $W$ and $V$ defined as follows:
\begin{eqnarray}
U(r) & = & \lb 1-\frac{\ln r}{\ln r_s}\rb^{1+\omega},\\
W(r) & = & \lb 1-\frac{\ln r}{\ln r_s}\rb^{1-\omega},\\
V(r) & = & \gamma^2\lb 1-\frac{\ln r}{\ln r_s}\rb^{\frac{1}{2}(\omega^2-1)}\exp\lp-\frac{\ln^2 r}{\ln r_s}\rp.
\end{eqnarray}
This metric presents a curvature singularity placed at finite proper distance $r_s=(8\pi F^2)^{-1}$, where $F$ is the vacuum expectation value of the scalar field whose symmetry breaking leads to the global string. The parameter  $\omega$ is related to the mass excess (per unit length and as a fraction of $M_{P}^2$, being $M_P$ the Planck mass) resulting from the existence of a time-like current along the string. The constant $\gamma$ can be determined by matching the metric (\ref{ckw}) with the metric inside he core; the radius of the core  is $r_{\mathrm {core}}\approx 1$  (see Ref. \cite{coka}). We shall assume $0<\omega<1$; in the limit $\omega\to 0$ the case studied in Ref. \cite{ei2} is recovered.

The extension of our analysis to this metric is slightly more complicated than the preceding one. It requires some care  with the preliminary study of the original manifold, because of the existence of a
curvature singularity at finite $r$ and because the flare-out  condition is not fulfilled everywhere for arbitrary values
of the parameters. However, the steps leading to an equation  allowing for a qualitative understanding of the dynamical
evolution turn out to be straightforward. The area per unit of $z$ coordinate, for a given value of the radius, is determined by the product of $W$, $V$ and $r^2$. Taking the corresponding derivative with respect to $r$, it is easy to show that the flare-out condition required for the possible existence of the associated thin-shell wormhole can be satisfied. Indeed, we find that the flare-out condition is  fulfilled as long as $a<r_s\exp{\lp-|\omega
-1|\sqrt{(\ln r_s)}/2\rp}$ is selected as the wormhole throat radius. Applying the cut and paste procedure, the components of the extrinsic curvature now turn  to be
\be
{K_\tau^\tau}^{\pm}  = \mp\frac{1}{2}\sqrt{\frac{V(a)}{1+V(a)\dot a ^2}}\left[2\ddot a+\dot a^2\left(\frac{U'(a)}{U(a)}+\frac{V'(a)}{V(a)}\right)+\frac{U'(a)}{U(a)V(a)}\right],
\label{e18a}
\ee
\begin{equation}
{K_\theta^\theta}^{\pm} = \pm \frac{1}{2}\sqrt{\frac{1+V(a)\dot a ^2}{V(a)}}\left( \frac{2}{a}+\frac{V'(a)}{V(a)}\right),
\label{e18b}
\end{equation}
and
\be
{K_z^z}^{\pm} = \pm\frac{1}{2} \sqrt{\frac{1+V(a)\dot a ^2}{V(a)}}\,\frac{W'(a)}{W(a)}. 
\label{e18c}
\ee
The resulting expressions for the energy density and pressures are
\be
  \sigma = -\frac{1}{8\pi}\sqrt{\frac{1+V(a)\dot a ^2}{V(a)}}\left(\frac{2}{a}+ \frac{W'(a)}{W(a)}+\frac{V'(a)}{V(a)}\right),
\ee
\be
p_{\theta} = \frac{1}{8\pi}\sqrt{\frac{V(a)}{1+V(a)\dot a ^2}}\left[2\ddot a+\dot a^2\left(\frac{U'(a)}{U(a)}+\frac{V'(a)}{V(a)}+\frac{W'(a)}{W(a)}\right)+\frac{1}{V(a)}\left(\frac{U'(a)}{U(a)}+\frac{W'(a)}{W(a)}\right)\right],
\label{e112}
\ee
\be
p_{z} = \frac{1}{8\pi}\sqrt{\frac{V(a)}{1+V(a)\dot a ^2}}\left[2\ddot a+\dot a^2\left(\frac{U'(a)}{U(a)}+2\frac{V'(a)}{V(a)}+\frac{2}{a}\right)+\frac{1}{V(a)}\left(\frac{U'(a)}{U(a)}+\frac{V'(a)}{V(a)}+\frac{2}{a}\right)\right].
\label{e122}
\ee
In the static case $\dot a=\ddot a=0$ the pressures are related with the energy density by the equations of state
\be
p_z=-\sigma\frac{U'(a)V(a)W(a)a+U(a)V'(a)W(a)a+2U(a)W(a)V(a)}{U(a)V(a)W'(a)a+U(a)V'(a)W(a)a+2U(a)W(a)V(a)}.
\ee
\be
p_\theta=-\sigma\frac{U'(a)V(a)W(a)a+U(a)V(a)W'(a)a}{U(a)V(a)W'(a)a+U(a)V'(a)W(a)a+2U(a)W(a)V(a)}.
\ee
If we propose, as before, that the static equations of state remain valid for the dynamic case (as long as the same hypothesis above are satisfied) then, after some algebraic manipulations,  we obtain the equation of motion
\be
2V(a)\ddot a+V'(a)\dot a^2=0
\ee 
The form of the equation of motion is the same of that obtained before, and the solution for the throat velocity is then
\be
\dot a(\tau)=\dot a_0\sqrt{\frac{V(a_0)}{V(a)}}.
\ee 
Therefore, after a small velocity perturbation  the wormhole throat undergoes a monotonous evolution. In other words, no oscillatory behaviour exists after perturbing an initially static configuration. Within the range $r_-<r<r_+$ with $r_\pm=\sqrt{r_s}\exp\lp\pm\sqrt{\ln r_s(\ln r_s+\omega^2-1)}/2\rp$ the function $V$ decreases  with $r$. For $0<\omega<1$, the radius $r_-$ turns to be smaller than the core radius, while $r_+$ results greater than the singularity radius. Thus for such values  of $\omega$ we have $V'<0$ for any physically meaningful  radius beyond the core and satisfying the flare-out condition. Consequently, the throat contracts decelerating if $\dot a<0$, while in the case of  of an  initial velocity pointing outwards the result is an accelerated expansion (as before,  the low velocity assumption is eventually no more fulfilled).

\section{Conclusion}
Summarizing, for both extensions of previous analysis within the framework of thin-shell wormholes associated to cosmic strings, we find that no oscillatory solutions exist for the class of equations of state adopted. A small inicial velocity pointing to the axis of symmetry leads to a decelerated contraction of the wormhole throat, while an initial slow expansion is accelerated.  Our results, thus, provide more  evidence supporting the recent conjecture that this class of wormholes would not be stable under small velocity perturbations, as long as the cylindrical symmetry and the static equations of state for matter on the shell are preserved.

\end{document}